\documentstyle[prl,aps,preprint]{revtex}

\begin{document}
%\draft
\title{On closed rotating worlds}

\author{Thoralf Chrobok, Yuri N.\ Obukhov\footnote{On leave from: Department
of Theoretical Physics, Moscow State University, 117234 Moscow,
Russia}, and Mike Scherfner\footnote{FB Mathematik, Technical
University of Berlin, Str. d. 17. Juni 136, D-10623 Berlin}}
\address
{Institute of Theoretical Physics, Technical University of Berlin,
Hardenbergstr. 36, D-10623 Berlin, Germany}

\maketitle
%\bigskip
% \noindent{\it file nine3.tex, 05 Feb 2001}
%\bigskip

\begin{abstract}
A new solution for the stationary closed world with rigid rotation is
obtained for the spinning fluid source. It is found that the spin and
the vorticity are locally balanced. This model qualitatively shows that 
the local rotation of the cosmological matter can be indeed related to
the global cosmic vorticity, provided the total angular momentum of 
the closed world is vanishing.
\end{abstract}

%\bigskip\bigskip
\pacs{PACS no.: 04.20.Cv; 04.20.Jb; 98.80.Hw}
%\bigskip

\section{Introduction}

The study of rotating cosmological models is of interest both
mathematically and physically (some of the previous work on this
subject is overviewed in \cite{coll,rotrev}). The first stationary
metric with vorticity \cite{goe1,goe2} described an open world filled
with dust matter plus the cosmological term. For many years and for
the numerous researchers, this model served a kind of theoretical
laboratory for the investigation of the specific rotational effects.
Besides, it stimulated the study of the role of Mach's principle in
the general relativity theory. In this respect, it is worthwhile to
mention the papers of Ozsv\'ath and Sch\"ucking \cite{osch} which
made a major contribution by providing the true anti-Machian model.
The latter describes a {\it spatially closed} world, in contrast to
the open universe of G\"odel.

One can distinguish the two types of rotation in cosmology. The first
one, which can be called geometrical (or spacetime) rotation, is
determined by the kinematics of the curved spacetime itself,
irrespectively of the matter contents of a model. Technically, its
properties are described by a formal comoving observer interpreting
the cosmic time coordinate line as an own world-line. The second type
of rotation arises when the cosmological matter itself has a proper
motion such that the corresponding fluid flow is characterized by the
nontrivial vorticity (which can be called material rotation).

In the original G\"odel's model \cite{goe1}, these two rotations
coincide because the matter is assumed comoving, however this is not
true in general \cite{panov}. A typical example is provided in
\cite{gron} where the matter flow has nontrivial vorticity, whereas
the spacetime (described by the Robertson-Walker metric) is
non-rotating. Another example is given by the anti-Machian closed
world model \cite{osch} in which the material and geometrical
rotations are different (moreover, the matter flow has a nontrivial
shear unlike the spacetime geometry itself).

In the study of the rotating cosmological models, it seems natural to
consider the spin (intrinsic angular momentum) as an important
physical property of the matter source. After all, the elements of
the ``cosmic fluid'' are either galaxies or clusters of galaxies (at
modern epoch) or the high energy elementary particles (at early
epochs) all of which have spin. Actually, that was one of the basic
motivations for Whittaker \cite{whit} to ask how can the largest
physical system (universe) be non-rotating when all the other known
systems, from particles to galaxies, have rotation.

In the present paper, we will demonstrate that the spinning cosmic
matter gives rise to a true anti-Machian rotating closed world. The
resulting model is, in a certain sense, a ``relative'' of the old
stationary model of Ozsv\'ath and Sch\"ucking \cite{osch}. Namely,
both models belong to the type-IX Bianchi spatially homogeneous
spacetimes. There is however, an essential difference in that the
material rotation now coincides with the spacetime rotation (i.e.,
there is no proper motion of matter) and the shear is absent, in
contrast to \cite{osch}. It seems worthwhile to note that our new 
model is stationary and thus, like the Ozsv\'ath and Sch\"ucking 
world \cite{osch}, it does not refer directly to the real expanding
universe.

\section{Spinning matter}

We will describe the cosmic matter with spin by means of the
Weyssenhoff model. It was proposed in flat spacetime in \cite{wey},
and the modern treatment of the curved spacetime variational theory
of spinning fluid is presented in \cite{ok}.

The ideal Weyssenhoff fluid is a continuous medium, the elements of
which are characterized by the intrinsic angular momentum (spin)
proportional to the volume element. The spin density is described by
the second rank skew-symmetric tensor
\begin{equation}
S^{ij} = -\,S^{ji},
\end{equation}
the spatial components of which coincide in the rest frame with the
density of the 3-vector of spin of a matter element. This is provided
by imposing the covariant constraint, the so called Frenkel
condition:
\begin{equation}
S_{ij}\,u^{i} = 0,\label{3}
\end{equation}
where $u^{i}$ is the four-velocity of a fluid element.

The (metric) energy-momentum tensor of the Weyssenhoff fluid reads:
\begin{equation}
T_{ij} = \varepsilon\,u_{i}u_{j} - p\,h_{ij} - 2(g^{kl} +
u^{k}u^{l})\,\nabla_{k}(u_{(i}S_{j)l}),\label{4}
\end{equation}
where $\varepsilon$ is the energy density, $p$ is the pressure, and
$h_{ij} = g_{ij} - u_{i}u_{j}$ is the standard projector (with the
velocity normalized by $u_{i}u^{i} = 1$).
The energy-momentum tensor (\ref{4}) can be written in a different
form of a general non-ideal fluid with ``the energy flux" $q^{i}$ and
``anisotropic pressure" $\pi_{ij}$ as
\begin{equation}
T_{ij}= \varepsilon_{\rm eff}\,u_{i}u_{j} - p_{\rm eff}\,h_{ij} + 2
u_{(i}q_{j)} + \pi_{ij},
\end{equation}
where as usually $u^{i}q_{i}=0, \pi^{i}{}_{i}=0, \pi_{ij}u^{i} =0$,
and the newly introduced quantities are defined by the spin:
\begin{eqnarray}
\varepsilon_{\rm eff} &=& \varepsilon - 2\omega_{kl}S^{kl}, \qquad
p_{\rm eff} = p - {2\over 3}\omega_{kl}S^{kl},\\ q_{i} &=&
-\,(h^{j}{}_{i}\nabla_{l}\,S_{j}{}^l + a^{l}\,S_{li}),\\ \pi_{ij} &=&
2\omega_{(i}{\,}^k\,S_{j)k} - 2\sigma_{(i}{\,}^k\,S_{j)k} - {\frac 2
3}\,h_{ij}\,\omega_{kl}\,S^{kl}.
\end{eqnarray}

The dynamics of spinning fluid is described by the rotational and
translational equations of motion. The rotational equations give the
motion of spin:
\begin{equation}
\nabla_i(u^i S_{kl}) = u_k u^j\nabla_i (u^i S_{jl}) - u_l u^j
\nabla_i(u^i S_{jk}) = u_k\,S_{lj}\,a^j - u_l\,S_{kj}\,a^j.\label{5}
\end{equation}
Hereafter $ a^{j} = u^{i}\nabla_{i}u^{j}$ is the acceleration field.
The translational equations of motion are, as usual, the consequence
of the energy-momentum conservation law. From (\ref{4}) one finds,
with the help of (\ref{3}),(\ref{5}) and the Ricci identity:
\begin{eqnarray}
\nabla_{i} T^{i}{}_{j} &=& u_{j}(u^{i}\nabla_{i}\varepsilon +
\varepsilon\nabla_{i}u^{i} +  p\nabla_{i}u^{i}) -
h^{i}{}_{j}\nabla_{i}\,p \nonumber\\ && +\,(p + \varepsilon)\,a_{j} +
2 u^{i}\,S_{jl}\nabla_{i}a^{l} + R_{klij}\,u^{i}\,S^{kl} =
0.\label{6}
\end{eqnarray}
Projections of (\ref{6}) on $u^{i}$ and on the orthogonal directions read
\begin{eqnarray}
(p + \varepsilon)\nabla_i u^i + u^i\nabla_i\varepsilon &=&
0,\label{7}\\ (p + \varepsilon)a_{j} - h^{i}{}_{j}\nabla_{i}\,p + 2
S_{ji}u^{l}\nabla_{l}a^{i} + R_{klij}\,u^{i}\,S^{kl} &=& 0.\label{8}
\end{eqnarray}

\section{Geometry}

We will study the stationary spacetime model belonging to the wide
class of rotating spatially homogeneous models \cite{rotrev}
\begin{equation}
ds^2 = dt^2 - 2\,R\,n_{a}dx^{a}dt -
R^2\,\gamma_{ab}\,dx^{a}dx^{b}.\label{met0}
\end{equation}
Hereafter the indices $a,b,c = 1,2,3$ label the spatial coordinates,
$R = R(t)$ is the scale factor, and
\begin{equation}
n_{a}=\nu_A\,e_a^{A},\qquad
\gamma_{ab}=\beta_{AB}\,e_a^{A}e_b^{B}.\label{ng}
\end{equation}
Here $\nu_{A},\beta_{AB}$ are constant coefficients ($A,B = 1,2,3$),
while
\begin{equation}
e^{A}= e_{a}^{A}(x)\,dx^a \label{ea}
\end{equation}
are the invariant $1$--forms with respect to the action of a
three-parameter group of motion which is admitted by the space-time
(\ref{met0}). We assume that this group acts simply-transitively on
the spatial ($t=const$) hypersurfaces. It is well known that there
exist 9 types of such manifolds, distinguished by the Killing vectors
$\xi_{A}$ and their commutators
$[\xi_{A},\xi_{B}]=f^{C}{}_{AB}\,\xi_{C}$. Invariant forms (\ref{ea})
solve the Lie equations ${\cal L}_{\xi_{B}}e^{A}=0$ for each Bianchi
type, so that models (\ref{met0}) are spatially homogeneous.

More specifically, we will confine our attention to the case of the
{\it closed world} belonging to the Bianchi type IX. Then, denoting
the spatial coordinates $x=x^1, y=x^2, z=x^3$, one has explicitly:
\begin{equation}
e^1 = \cos y\,\cos z\,dx - \sin z\,dy,\quad e^2 = \cos y\,\sin z\,dx
+ \cos z\,dy,\quad e^3 = -\,\sin y\,dx + dz,\label{eA}
\end{equation}
which satisfy the structure equations
\begin{equation}
de^A = f^A{}_{BC}\,e^B\wedge e^C,\quad {\rm with}\quad f^1{}_{23} =
f^2{}_{31} = f^3{}_{12} = 1.
\end{equation}

We now write the ansatz for the line element (\ref{met0})
\begin{equation}
ds^2 = g_{\alpha\beta}\,\vartheta^\alpha\,\vartheta^\beta, \qquad
 g_{\alpha\beta} = {\rm diag}(1, -1, -1, -1),
\end{equation}
in terms of the orthonormal coframe 1-forms $\vartheta^\alpha$:
\begin{equation}
\vartheta^{\widehat{0}} = dt - R\,\nu_A\,e^A,\
\vartheta^{\widehat{1}} = R\,k_1\,e^1,\ \vartheta^{\widehat{2}} =
R\,k_2\,e^2,\ \vartheta^{\widehat{3}} = R\,k_3\,e^3.\label{var}
\end{equation}
Here, $k_1, k_2, k_3$ are positive constant parameters. The Greek
indices $\alpha,\beta,\dots = 0,1,2,3$ hereafter label the objects
with respect to the orthonormal frame; the hats over indices denote
the separate frame components of these objects.

\section{Closed rotating world}

In order to finalize the formulation of the problem, we have to
specify the assumptions about the cosmic fluid source. As we already
mentioned, we are interested in the comoving case, when the mean
velocity is
\begin{equation}
u^i = \delta^i_0.\label{U}
\end{equation}
Hence, with respect to the orthonormal frame, $u^\alpha = (1, 0, 0,
0)$, and from the Frenkel condition (\ref{3}) we see that the spin
density can have only spatial nontrivial components:
\begin{equation}
S_{\widehat{2}\widehat{3}},\qquad S_{\widehat{3}\widehat{1}},\qquad
S_{\widehat{1}\widehat{2}}.\label{Svec}
\end{equation}

We are looking for the stationary world, so all the time dependence
is absent, and thus the scale factor is fixed at some constant value,
$R = R_0$, whereas the components of spin (\ref{Svec}) are also
constant.

The gravitational field equations read, with $\kappa$ the Einstein
gravitational constant and cosmological constant $\Lambda$ (with
respect to the orthonormal local frame):
\begin{equation}
R_{\alpha\beta} - {1\over 2}\,R\,g_{\alpha\beta} =
\kappa\,T_{\alpha\beta} + \Lambda\,g_{\alpha\beta}.\label{ein}
\end{equation}
Substituting (\ref{var})-(\ref{Svec}) into (\ref{4}) and (\ref{ein}),
after some algebra one finds from the {\it off-diagonal} components
of the Einstein equations:
\begin{equation}
\kappa\,S_{\widehat{2}\widehat{3}} = {\frac
{\nu_1}{2R_0k_2k_3}},\qquad \kappa\,S_{\widehat{3}\widehat{1}} =
{\frac {\nu_2}{2R_0k_1k_3}},\qquad \kappa\,S_{\widehat{1}\widehat{2}}
= {\frac {\nu_3}{2R_0k_1k_2}}.\label{Ssol}
\end{equation}
Using these in (\ref{5}), one verifies that the rotational equations
of motion are satisfied automatically. Substituting (\ref{Ssol}) into
the {\it diagonal} components of (\ref{ein}), one  finds that the
latter are consistent only when all the three parameters are equal:
\begin{equation}
k_1 = k_2 = k_3 =: k.
\end{equation}
Then, the Einstein equations yield for the energy density and the
pressure
\begin{eqnarray}
\kappa\,\varepsilon + \Lambda &=& {\frac {3k^2 - \nu_1^2 - \nu_2^2 -
\nu_3^2}{4R_0^2k^4}},\\ \kappa\,p - \Lambda &=& {\frac {- k^2 -
\nu_1^2 - \nu_2^2 - \nu_3^2}{4R_0^2k^4}}.
\end{eqnarray}

For the {\it dust} equation of state $p = 0$ (incoherent matter), we
thus finally find:
\begin{equation}\label{Esol}
\varepsilon = {\frac {2(k^2 - \nu_1^2 - \nu_2^2 - \nu_3^2)}{4\kappa
R_0^2k^4}}, \qquad \Lambda = {\frac {k^2 + \nu_1^2 + \nu_2^2 +
\nu_3^2}{4\kappa R_0^2k^4}}.
\end{equation}

\section{Discussion and conclusion}

It is straightforward to obtain the kinematical quantities which
describe the spacetime geometry. A direct calculation of the
vorticity, $\omega_{\mu\nu} = h^{\alpha}{}_{\mu}h^{\beta}{}_{\nu}
\nabla_{[\alpha}u_{\beta]}$, shear,
$\sigma_{\mu\nu}=h^{\alpha}{}_{\mu}
h^{\beta}{}_{\nu}\nabla_{(\alpha}u_{\beta)} - {1\over 3}\,h_{\mu\nu}
\nabla_{\lambda}u^{\lambda}$, and the volume expansion $\theta=
\nabla_{\lambda}u^{\lambda}$, yields:
\begin{eqnarray}
&& \sigma_{\mu\nu} = 0,\qquad a^\alpha = 0,\qquad \theta = 0,\\ &&
\omega_{\widehat{2}\widehat{3}} = -\,{\frac {\nu_1}{2R_0k^2}},\qquad
\omega_{\widehat{3}\widehat{1}} = -\,{\frac {\nu_2}{2R_0k^2}},\qquad
\omega_{\widehat{1}\widehat{2}} = -\,{\frac
{\nu_3}{2R_0k^2}}.\label{kin}
\end{eqnarray}
Comparing with (\ref{Ssol}), we find
\begin{equation}
\kappa\,S_{\alpha\beta} = -\,\omega_{\alpha\beta}.\label{SO}
\end{equation}
This is quite a remarkable relation as it indeed supports the idea
which was put forward by Whittaker \cite{whit}: the rotation of the
local matter elements, i.e. spin $S_{\alpha\beta}$, is derived from
the overall global rotation of the universe, that is
$\omega_{\alpha\beta}$. Moreover this result is in complete agreement
with the work of King \cite{king} on closed rotating Bianchi-IX
spacetimes. The rotation of the local matter elements are balanced by
the rotation of the spacetime, so that there is no net angular
momentum. This can be made more explicit by recasting (\ref{SO}) into
the form of the equation describing the vanishing of the total
(orbital plus spin) angular momentum of the universe,
\begin{equation}
\kappa\,S_{\alpha\beta} + \omega_{\alpha\beta} =0.
\end{equation}
Thus, in the sense of King, our solution is not a counterexample to
Mach's principle. We however have a more direct and transparent
mechanism of the balance, without the need of the gravitational-wave
interpretation developed in \cite{king}. Note that our results have
been obtained in the framework of Einstein's general relativity 
theory. Similar observations were also reported for the cosmic 
spinning string solutions in the Einstein-Cartan theory of 
gravity \cite{soleng}.

Another interesting result is that, given the radius of the universe
$R_0$, the magnitude of the angular velocity $\omega^2 :={\frac 1 2}
\omega_{\alpha\beta}\omega^{\alpha\beta} = (\nu_1^2 + \nu_2^2 +
\nu_3^2)/(4R_0^2k^4)$ cannot be arbitrarily large in the closed
world. As we see from (\ref{Esol}), the energy density is only
positive when
\begin{equation}
|\omega | < {\frac 1 {2R_0k}}.
\end{equation}
Recall that in the open G\"odel cosmos $\kappa\,\varepsilon =
2\omega^2$, and thus the angular velocity could be arbitrary.

\section{Acknowledgments}
This work was supported by the Deutsche Forschungsgemeinschaft with
the grants HE 1922/5-1 and 436 RUS 17/72/00.

%\begin{thebibliography}{999}         %for Latex

\end{document}